\documentclass[amsart,11pt]{revtex4}

\usepackage{amsmath,amsfonts,amssymb,bm,graphicx,times}
\usepackage[usenames]{color}

\begin{document}

\title{Noise-Powered Probabilistic Concentration of Phase Information}

\author{Mario A. Usuga$^{*1,2}$, Christian R. M\"uller$^{*1,3}$, Christoffer Wittmann $^{1,3}$, Petr Marek $^{4}$, Radim Filip $^{4}$, Christoph Marquardt $^{1,3}$, Gerd Leuchs$^{1,3}$ and Ulrik L. Andersen $^{2}$}

\affiliation{
\\$^{1}$ Max Planck Institute for the Science of Light, Guenther-Scharowsky-Str. 1, 91058 Erlangen 
\\$^{2}$ Department of Physics, Technical University of Denmark, 2800 Kongens Lyngby, Denmark
\\$^{3}$ Institute for Optics, Information and Photonics, University Erlangen-Nuremberg, Staudtstr. 7/B2, 91058 Erlangen
\\$^{4}$ Department of Optics, Palack\'{y} University 17,  listopadu 50, 772 07 Olomouc, Czech Republic
\\$^{*}$These authors contributed equally to this work
 }

\maketitle

Phase insensitive optical amplification of an unknown quantum state is known to be a fundamentally noisy operation that inevitably adds noise to the amplified state~\cite{louisell,haus_mullen_1962,kimble,caves,josse}. However, this fundamental noise penalty in amplification can be circumvented by resorting to a probabilistic scheme as recently proposed and demonstrated in refs~\cite{ralph_lund_2009, pryde, grangier_ONA}. These amplifiers are based on {\it highly non-classical} resources in a complex interferometer. Here we demonstrate a probabilistic quantum amplifier beating the fundamental quantum limit utilizing a {\it thermal noise} source and a photon number subtraction scheme~\cite{marek_filip_th}. The experiment shows, surprisingly, that the addition of incoherent noise leads to a noiselessly amplified output state with a phase uncertainty below the uncertainty of the state prior to amplification. This amplifier might become a valuable quantum tool in future quantum metrological schemes and quantum communication protocols.

Besides being the subject of a fundamental dicussion going back to Dirac~\cite{Dirac}, the measurement of phase is at the heart of many quantum metrological and quantum informational applications such as gravitational wave detection, global positioning, clock syncronization, quantum computing and quantum key distribution. In many of these applications the phase is most often imprinted onto a coherent state of light and subsequently estimated using an interferometric measurement scheme. Such a phase estimation process~\cite{shapiro} is however hampered by the fundamental quantum noise of the coherent state which plays an increasingly devastating role as the excitation of the coherent state becomes smaller. Small coherent state excitations and associated large phase uncertainties are typical in real systems such as long distance coherent state communication and lossy interferometry.   

To reduce the phase uncertainty and thus concentrate the phase information, the state must be amplified noiselessly. This can be done probabilistically using either a highly complicated interferometric setup of single photon sources~\cite{ralph_lund_2009, pryde, grangier_ONA}, a sophisticated sequence of photon addition and subtraction schemes~\cite{marek_filip_th, fiurasek} or a very strong cross-Kerr nonlinearity~\cite{menzies}. However, as we show in this letter, it is possible to amplify the phase information noiselessly without the use of any non-classical resources or any strong parametric interactions. Remarkably, the supply of energy in our amplifier is simply a thermal light source. 

\begin{figure}
\includegraphics[width=15cm]{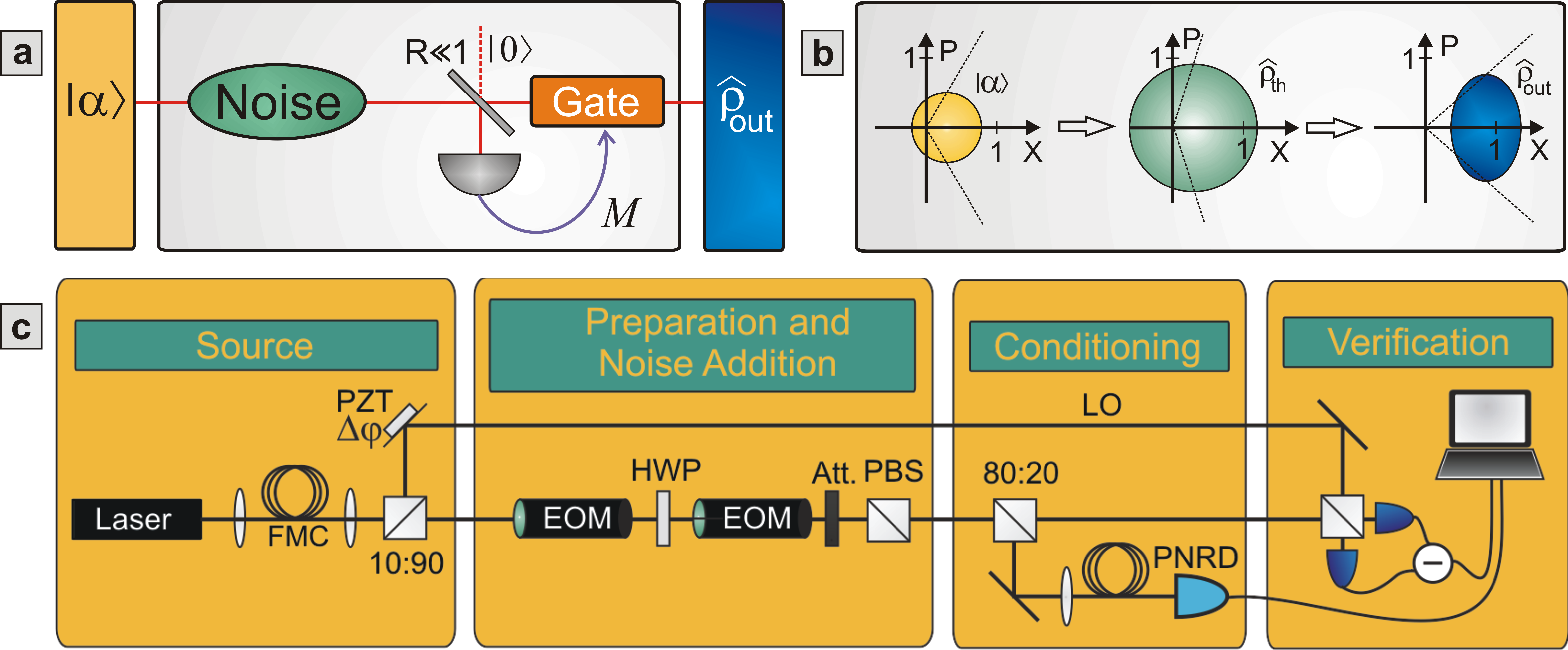}
\caption{a) Principal setup: Noise is incoherently added on an input state. Subsequently a part of the beam is tapped off and measured by a photon number resolving detector (PNRD). Results of that measurement that surpass a specified threshold M, herald the output state. b) Principal operation in phase space: A coherent state (yellow) serves as the input. The dashed line indicates the phase variance. Thermal noise is added to this input state resulting in a displaced thermal state (green). The output state (blue) is reshaped and the resulting phase variance is reduced compared to the input state. c) Experimental setup: An external cavity diode laser with fiber mode cleaning (FMC) acts as the source for the experiment and is split into a local oscillator (LO) and an auxiliary oscillator. The signal is prepared by a combination of two electro-optical modulators (EOM), a half-wave plate (HWP) and attenuation (Att.). A polarizing beam splitter removes the auxiliary oscillator. Part of the signal is tapped by a 80:20 beam splitter and coupled via a multimode fibre into the photo number resolving detector. This measurement is conditioning the output state, which is characterized by a homodyne measurement. The phase of the LO is controlled by a piezoelectric transducer (PZT).} 
\label{fig1}
\end{figure}

A schematic diagram of the probabilistic amplifier ~\cite{marek_filip_th} is shown in Fig.~\ref{fig1}a. It is solely based on phase insensitive noise addition and photon subtraction. 
To explain in simple terms why the addition of noise can help amplifying a coherent state, we consider the phase space pictures in Fig.~\ref{fig1}b. The addition of thermal noise induces random displacements to the coherent state, thus resulting in a Gaussian mixture of coherent states; some with excitations that are larger than the original excitation and some with smaller excitations. In the photon subtraction process, the coherent states with large excitations are probabilistically heralded, thereby rendering the state in a mixture consisting of the most excited coherent states from the original Gaussian mixture. As illustrated in Fig.~\ref{fig1}b, the resulting state is amplified and possess a reduced phase uncertainty.

The probabilistic photon subtraction procedure can be approximated by a largely asymmetric beam splitter combined with a photon number resolving detector, PNRD (see Fig.~\ref{fig1}a). A small portion of the displaced thermal state is directed to the photon counter and when a pre-specified number of photons is detected the transmitted state is heralded. Such an approach for photon number subtraction has also been employed for the generation of coherent state superpositions~\cite{ourjoumtsev,nielsen}. However, in contrast to previous implementations that were limited to the demonstration of two-photon subtraction~\cite{takahashi}, here we subtract up to four photons.

To elucidate the function of the amplifier, theoretically, we consider the amplification of a small amplitude ($|\alpha|\ll 1$) coherent state which can be approximately described in the two-dimensional Fock space: $|\alpha\rangle \approx |0\rangle+\alpha|1\rangle$.
Since the amplitude is small, the canonical phase variance of this state is to a very good approximation given by~\cite{wiseman}
\begin{equation}
V_C\approx\frac{1}{|\alpha|^2}
\label{cohVar}
\end{equation}
This variance represents the fundamental uncertainty in estimating the phase of the coherent state when a hypothetically ideal phase measurement is employed~\cite{armen}. The aim is to produce an amplified state with a phase variance reduced with respect to the coherent state variance in (\ref{cohVar}), thereby concentrating the phase information.
If a conventional phase insensitive amplifier is used to amplify the coherent state, the resulting variance is larger than (\ref{cohVar}) (see Apendix B).
On the other hand, if our amplifier is employed with weak Gaussian noise addition followed by single photon subtraction, the resulting state is~\cite{marek_filip_th}
\begin{equation}\label{rho_approx}
\hat\rho\approx\frac{1}{|\alpha|^2+N_\mathrm{th}+4|\alpha|^2N_\mathrm{th}}\left[ |\alpha|^2|0\rangle\langle 0|+N_\mathrm{th}(|0\rangle + 2\alpha|1\rangle)(\langle 0| + \langle 1| 2\alpha^*)\right]
\end{equation}
with the canonical phase variance
\begin{equation}
V_C^{amp} \approx \frac{1}{4|\alpha|^2}\left(1 + \frac{|\alpha|^2}{N_\mathrm{th}}\right) - 1. 
\end{equation}
where it is assumed that the average number of incoherently added photons is $N_\mathrm{th} \ll 1$. We quantify the performance of the amplifier by the normalized phase variance, $\Gamma = V_C^{amp}/V_C$ which is smaller than one for a noiseless operation. For the above approximative example, if $ |\alpha|^2\ll N_\mathrm{th}$, we find that the normalized variance approaches $\Gamma=1/4$. Another parameter that will be used to evaluate the amplifier is the gain $g=|\beta|/|\alpha|$ (where $\beta$ is the average amplitude of the output state), being $g=2$ for the above example. We therefore see that by simply adding a small amount of noise to the input state followed by single photon subtraction it is possible to create an output state with twice the amplitude and with a reduced phase variance.  

\begin{figure}
\includegraphics[width=10cm]{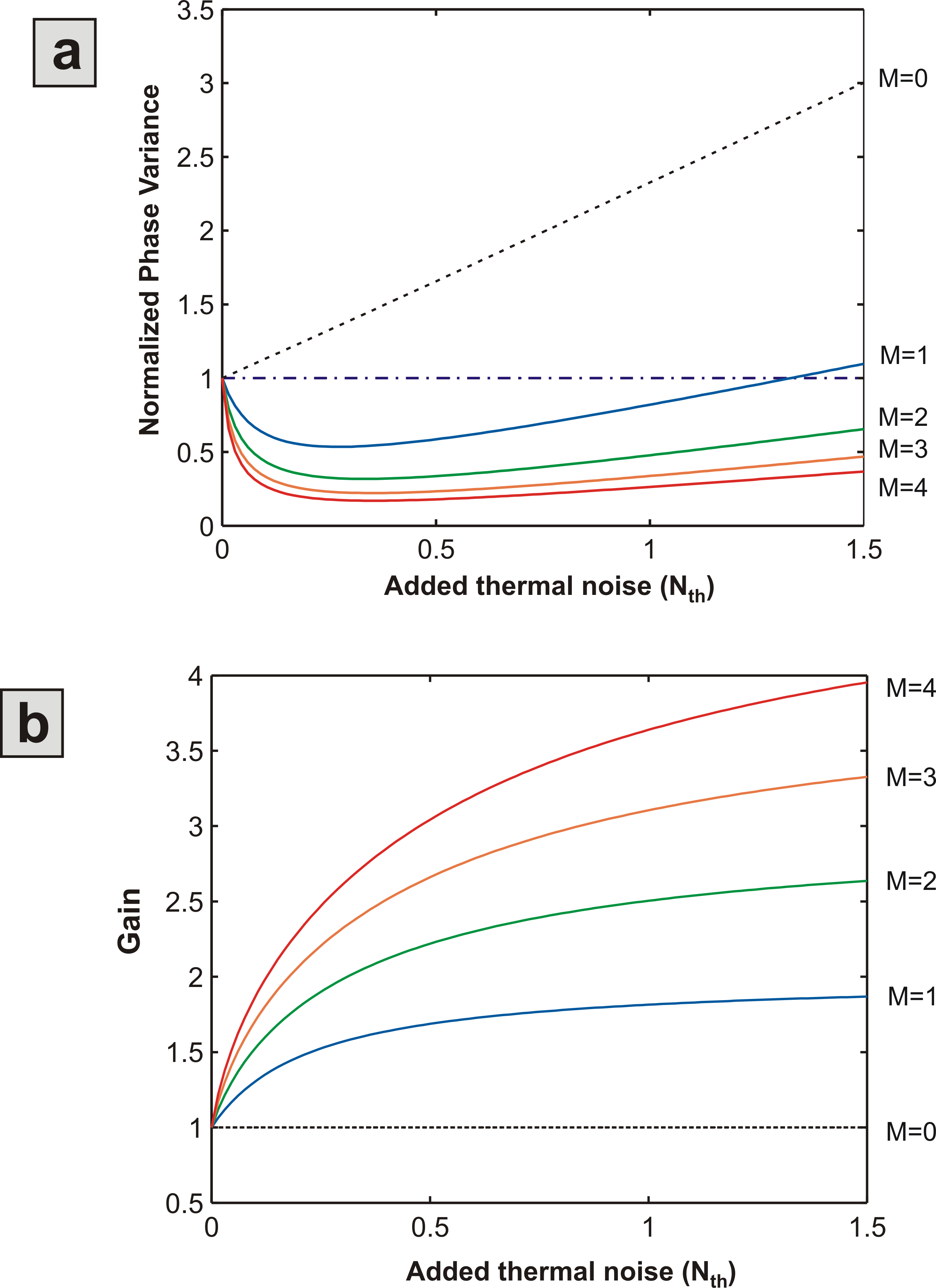}
\caption {\label{ExpSetup}Theoretical gain and normalized phase variance vs. the mean number of added thermal photons: a) The canonical variance normalized to the corresponding variance of the input coherent state. b) The gain generally grows with the number of photons in the added thermal noise and the threshold M of added thermal photons for an input coherent state of amplitude $\left|\alpha\right|=0.48$.  } 
\label{fig2}
\end{figure}

Based on a more general model (as presented in Apendix C), in Fig.~\ref{fig2} we plot the normalized phase variance and the gain as a function of the average number of added thermal photons, where $M$ denotes the threshold for the number of photon subtractions. The figures illustrate three interesting aspects: the phase noise reducing operation works even when the parameters go beyond the simple approximation considered above, the effect of the amplification improves with subtracting more photons, and the amount of noise must be relatively large for the amplification to work well. These aspects also follow the intuitive picture discussed above: Large noise addition will partly displace the coherent state in a radial direction in phase space with a correspondingly large magnitude, and the states with the largest amplitude (associated with strongly amplified states) are heralded by high photon number subtractions. We also note that the amount of added noise that minimizes the canonical variance depends solely on the magnitude of the input coherent state but not on its phase. The amplifier is thus capable of concentrating an unknown phase of a coherent state.

A laboratory implementation of the amplifier is depicted in Fig.~\ref{fig1}c. Our source is a grating stabilized CW diode laser operating at $809\,\mathrm{nm}$ with a coherence time of $1\,\mathrm{\mu s}$. The laser output is spatially cleaned in an optical fiber, and subsequently split to serve as a Local Oscillator (LO) for homodyne detection and as an auxiliary beam for state preparation. We use a pair of electro-optical modulators for the preparation of the displaced thermal state (corresponding to a coherent state with added thermal noise) in a polarization mode orthogonal to the polarization of the auxiliary mode (see Apendix A). The duration of the prepared pulses is $800\,\mathrm{ns}$. A portion (20\%) of the prepared state is tapped off in an asymmetric beam splitter and measured with an avalanche photo detector (APD) operating in an actively gated mode such that the dead time ($50\,\mathrm{ns}$) is much shorter than the pulse duration. This means that the APD can be used as a single photon counter provided that the mean number of photons in the detected pulse is very small. The transmitted part of the state is passed on to the homodyne detector where it interferes with a phase controlled LO. This provides quadrature measurements of the emerging states under any phase space angle. 
The measurement outcome is sent to the computer where it is postselected according to the result of the photon counting measurement. 
Based on the resulting data points we reconstruct the density matrix of the heralded state using a maximum likelihood algorithm~\cite{lvosky,hradil}. We correct the data for the inefficiency of the homodyne detector in order to reconstruct the actual input state and the amplified output state (see Apendix A).

\begin{figure}
 \includegraphics[width=12cm]{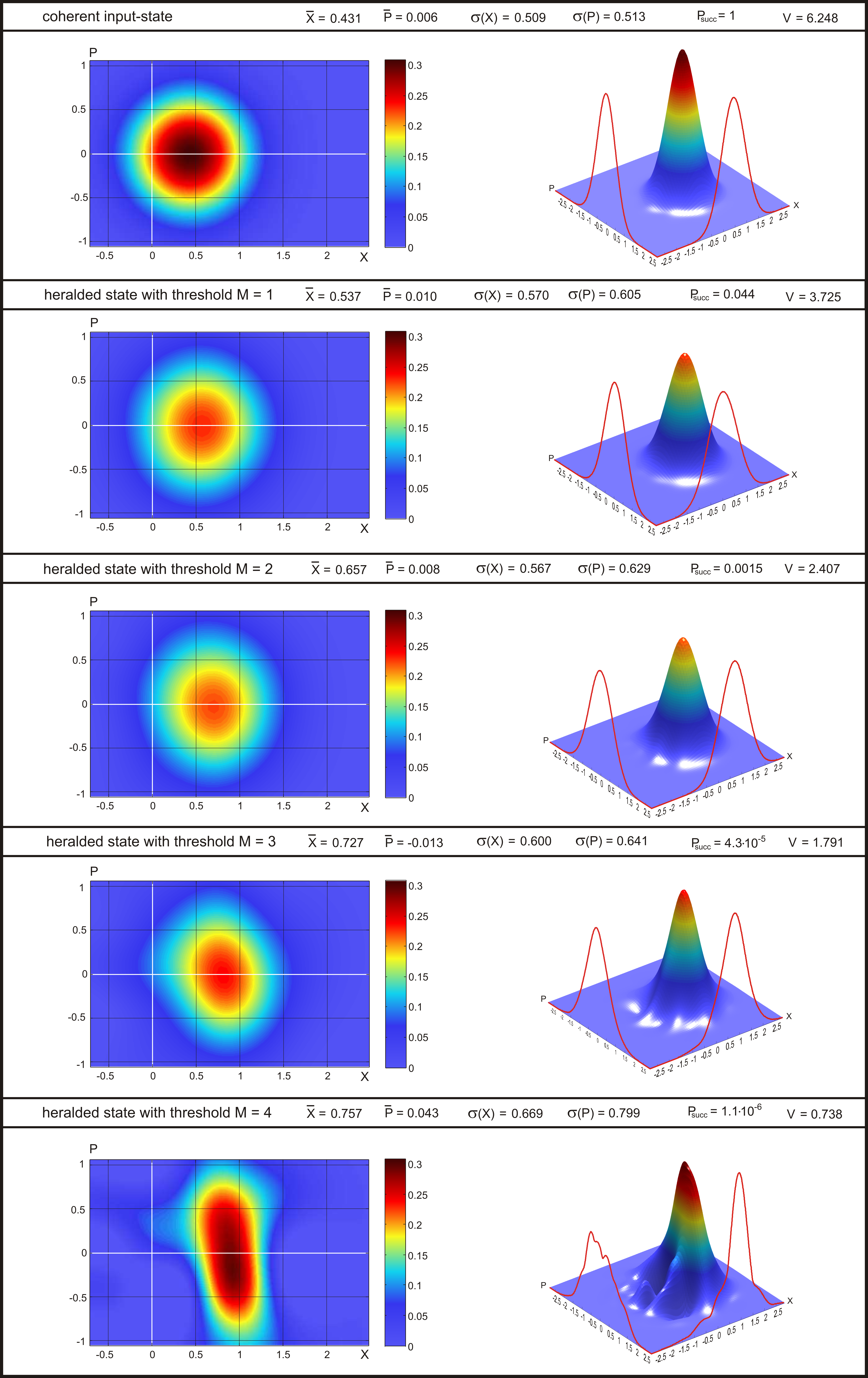}

\caption   {Reconstructed Wigner functions: Wigner functions that were reconstructed from experimental data for the input state and the heralded state for different thresholds M. For each experimental reconstruction, the mean values and standard deviations for the X and P quadratures are given with the corresponding measured success probability and canonical phase variance.} 
\label{fig3}
\end{figure}

\begin{figure}
\includegraphics[width=10cm]{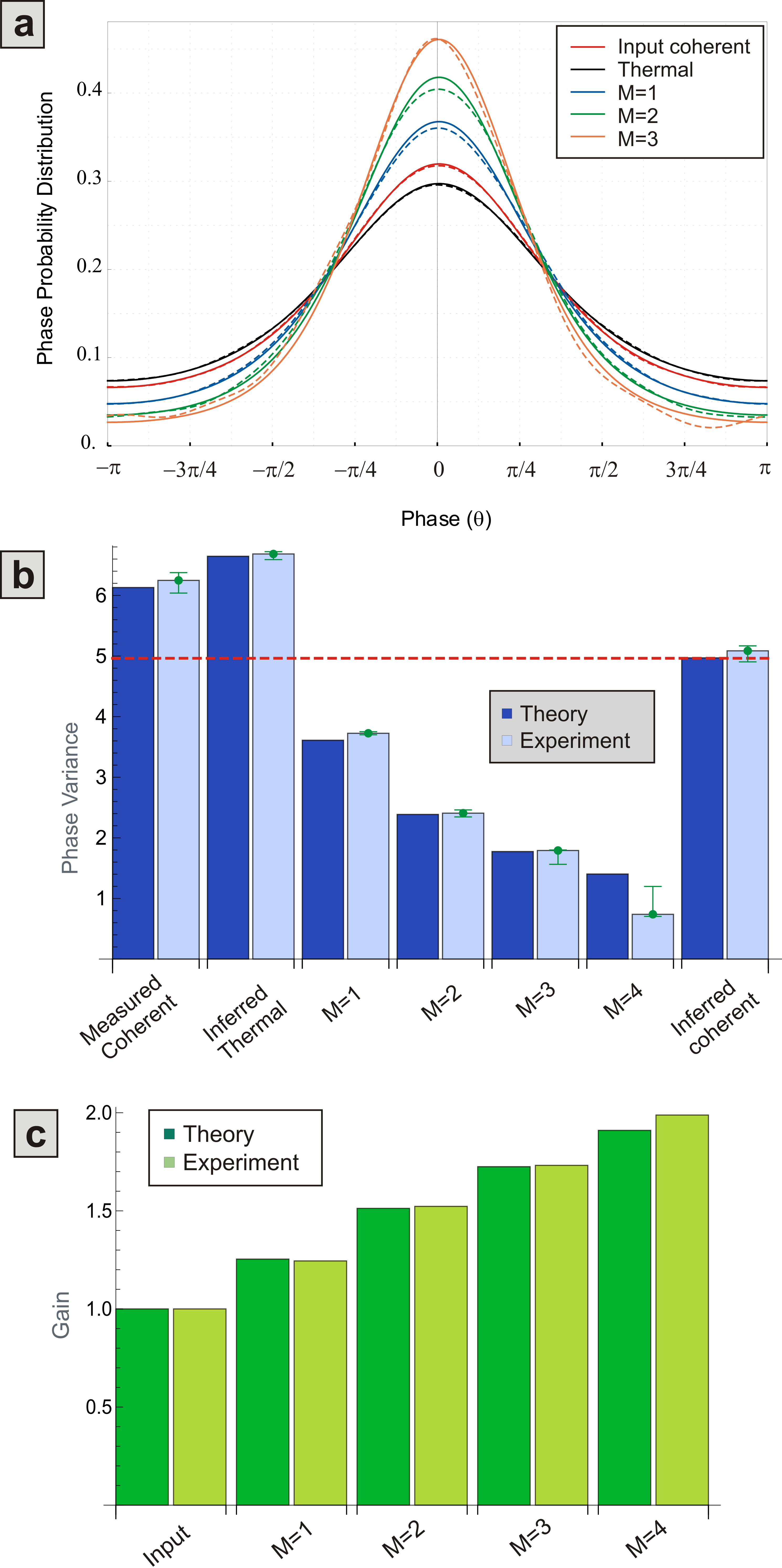}
\caption {a) Phase probability distribution function derived from the experimental data (solid lines) for the measured coherent, the thermal and the conditioned states. Corresponding theoretical functions (dashed lines) were calculated for states fitting to experimentally derived parameters. b) The canonical phase variance deduced from the experimental data (light blue) and corresponding theoretical values (dark blue) calculated for states fitting to the experimentally derived parameters. The inferred input coherent state serves as the reference value. c) Gain for the input coherent state for different thresholds M. The error bars in b) represent the statistical deviations over many different realizations of the experiment.\label{Theory}} 
\label{fig4}
\end{figure}

From the density matrices, we construct the Wigner functions for the input state and the amplified output states for different photon number subtractions as illustrated in Fig.~\ref{fig3}. Here we consider an input state excited along the amplitude quadrature axis with $\langle{X}\rangle=0.431$ thus $|\alpha|^2=0.186$ and a thermal noise addition corresponding to $\langle N_\mathrm{th}\rangle=0.15$. The amplification factor for this experiment is summarized in Fig.~\ref{fig4}c. We also reconstruct the phase distributions (see Apendix B) for different subtractions, the results of which are shown in Fig.~\ref{fig4}a. We clearly see that as the number of subtractions increases, the distribution becomes narrower, and thus the phase information is concentrated. These results are summarised in Fig.~\ref{fig4}b.

To optimise the performance of the amplifier - that is, to minimize the phase variance - the amount of added thermal noise should be chosen appropriately with respect to the input coherent state amplitude. Furthermore, we note that having detailed information about the input alphabet, the structure of the noisy displacements can be tailored accordingly, thereby drastically reducing the amount of energy used to drive the amplifier. For example if the input is a phase-covariant coherent state alphabet, the optimized structure of the noisy displacement is also phase covariant. Such tailoring of the displacements as well as applications of the amplifier will be interesting directions for future research. Finally, we note that the noise addition process can be also carried out with a linear amplifier. Such an approach will not only add thermal noise to the input state but will also displace it coherently in the preferred direction, thereby further concentrating the phase information.                 

In summary, we have reduced the phase uncertainty of a coherent state of light through noiseless probabilistic amplification. In contrast to previous approaches to noisefree amplification, the amplifier is neither based on an ample supply of nonclassical resources nor on strong parametric interactions, but solely on Gaussian noise addition and photon counting. Due to its pivotal properties such as simplicity and robustness, we expect that this approach to probabilistic noisefree amplification will be of interest to a large variety of experiments and protocols involving phase estimation such as quantum metrology and quantum communication.

\section*{APENDIX A}

The experimental setup in Fig.~\ref{fig1}c is described in the following. The laser source is monitored, to assure quantum-noise-limited signal states.
The states are prepared by two electro optical modulators (EOM) and a half wave plate (HWP). The modulators displace the signal state (S) using the orthogonally polarized auxiliary oscillator  mode (AO), which is relatively bright~\cite{wittmann}. After the modulators' calibration, we can displace the signal state to any coherent state with a maximum photon number corresponding to the mean photon number in the AO mode $n_\mathrm{max}=|\alpha_\mathrm{AO}|^2$. The signal mode can be chosen to be in an arbitrary mixed state, provided that the state's P-function is positive. We can therefore generate the displaced thermal state applying a suitable modulation sequence to both EOMs. The state is modeled with a finite set of (more than) $10^3$ coherent states, randomly picked from a 2D normal distribution. The random modulation sequence is varied and repeated throughout the measurement.

The tap beam is focused into the fibre coupled PNRD (for details see~\cite{banaszek, wittmann2}). The transmitted part of the beam is sent to a homodyne detector, which measures the signal with a continuously-scanned local oscillators' phase. The scanning frequency is chosen to be  $21\,\mathrm{mHz}$, leading to an effective phase drift of only $1.6\,\mathrm{mrad}$  within the modulation sequence. This value is negligible from an experimental point of view so that the LO's phase is considered constant within a single modulation sequence. The phase angle needed in the tomography was estimated with a series of phase calibration signals prepended to the modulation sequence. The main source of error in the setup is the drift of the modulators. To compensate for this drift, the calibration point was continuously adjusted.

As the homodyne detector (HD) is not a part of the phase concentration scheme itself but only implemented in the setup to prove the effect of the scheme, we do not want to take its imperfections into account in the analysis. We assume therefore perfect detection. The amplitude of the coherent input state $|\alpha|$ is then inferred from the measured mean photon number values in the imperfect PNRD and the ideal HD 
\begin{equation}
|\alpha|^2 = |\alpha_\mathrm{HD}|^2 + \frac{1}{\eta_\mathrm{PNRD}}|\alpha_\mathrm{PNRD}|^2
\end{equation}
where the PNRD's quantum efficiency was calibrated to be $\eta_\mathrm{PNRD} =0.63\pm3$\% using the overall quantum efficiency of the HD.
This procedure is preferable as it does not demand an accurate knowledge of the input coherent states amplitude, the splitting ratio and the losses in the HD.

\section*{APENDIX B}

Each measurement procedure devised to estimate a phase of a given quantum state can be characterized by a real semi-definite matrix $H$. The actual probability distribution of the estimated phase value can be obtained as $P(\theta) = \mbox{Tr}[\hat{\rho} \hat{F}(\theta)]$,
where $ \hat{F}(\theta) = 1/2\pi \sum_{m,n=0}^{\infty}\exp(i\theta(m-n))H_{mn}|m\rangle\langle n|$ ~\cite{wiseman}. To characterize the quality of phase encoding by a single value, we can use the Holevo phase variance ~\cite{holevo_var} $V = |\mu|^{-2} -1$, where $\mu = \langle \exp(i\theta)\rangle$. The fundamental limit of the phase estimation is obtained for the canonical measurement, in which the operator $\hat{F}(\theta)$ projects onto the idealized phase state $\sum_{n=0}^{\infty}e^{i\theta n}|n\rangle$ and $H \equiv 1$. The canonical phase variance is therefore given by

\begin{equation}\label{mu1}
    V_C = \left|\mbox{Tr}\left\{\sum_{n=0}^{\infty}|n\rangle\langle n+1|\hat{\rho}\right\}\right|^{-2}-1.
\end{equation}

\section*{APENDIX C}

In the following we theoretically describe the amplification process and explicitly calculate values of $\mu $ for various states. For a canonical measurement of an initial coherent state with amplitude $\alpha$, we find \cite{marek_filip_th}
\begin{equation}\label{mucoh}
    \mu = \mathrm{Tr}\left[\sum_{n=0}^{\infty}|n\rangle\langle n+1|\alpha\rangle\langle\alpha|\right]=
    e^{-|\alpha|^2}\alpha\sum_{n=0}^{\infty} \frac{|\alpha|^{2n}}{n!\sqrt{n+1}}
\end{equation}
It can be seen that the fundamental limit on phase estimation given by the canonical variance is fully given by the mean number of photons of the state. When the coherent state gets amplified by a parametric device with gain $G$, the result is a displaced thermal state state with $\alpha' = \sqrt{G}\alpha$, $N_\mathrm{th} = 2G -2$, and \cite{marek_filip_th}
\begin{equation}\label{}
    \mu = \frac{\alpha'}{\pi}\int_{0}^{\frac{2}{N_\mathrm{th}+2}}\frac{\exp(-x|\alpha'|^2)}{\sqrt{-\ln\left[1-\frac{2}{2-x N_\mathrm{th}}\right]}} dx,
\end{equation}
There is another way of obtaining $\mu$, which will be useful later. At its core stands the Glauber-Sudarshan representation \cite{Glauber}, which allows any quantum state to be expressed as
\begin{equation}\label{Pfun}
    \hat{\rho} = \int \Phi(\beta)|\beta\rangle\langle\beta| d^2\beta,
\end{equation}
where the $\Phi(\beta)$ is its P-function. For a coherent state it is a delta function $\delta(\beta-\alpha)$, while for a coherent state with thermal noise it is a Gaussian function
\begin{equation}\label{thermalcoh}
    \Phi(\beta) = \frac{1}{\pi N_\mathrm{th}} \exp\left(-\frac{|\beta-\alpha|^2}{N_\mathrm{th}}\right),
\end{equation}
where $N_\mathrm{th}$ denotes the mean number of thermal photons. Since $\mu$ is a linear functional of the density operator, we can, for any state whose P-function is known, obtain its value directly from (\ref{Pfun}) and (\ref{mucoh}). For a displaced thermal state with (\ref{thermalcoh}) we arrive at
\begin{equation}\label{}
    \mu = \sum_{n=0}^{\infty}\frac{1}{n!\sqrt{n+1}}\frac{e^{-\frac{|\alpha|^2}{N_\mathrm{th}+1}}}{N_\mathrm{th}+1} \mathcal{I}_n\left(\frac{\alpha}{N_\mathrm{th}+1},\frac{N_\mathrm{th}}{2(N_\mathrm{th}+1)}\right).
\end{equation}
Here we have defined the function $\mathcal{I}_n(A,B)$ of a positive integer $n$, a complex number $A = A_r + i A_i$ and a real number $B$, as
\begin{eqnarray}\label{Idef}
    \mathcal{I}_{n}(A,B) = \int (\beta_r + i\beta_i)(\beta_r^2 + \beta_i^2)^n \mathcal{G}_{\beta_r}(A_r,B)\mathcal{G}_{\beta_i}(A_i,B)d\beta_r d\beta_i \nonumber \\
    =\sum_{k=0}^{n} {n\choose k} [ \mathcal{M}_{2k+1}(A_r,B)\mathcal{M}_{2(n-k)}(A_i,B) + i \mathcal{M}_{2k}(A_r,B)\mathcal{M}_{2(n-k)+1}(A_i,B)],
\end{eqnarray}
where $\mathcal{G}_{\beta_.}(A,B) = \exp(-(\beta_. - A)^2/2B)/\sqrt{2 \pi B}$ is a Gaussian distribution of a variable $\beta_.$ with mean value $A$ and variance $B$, and $\mathcal{M}_k(A,B)$ stands for its $k$-th moment. The moments can be easily evaluated with the help of the moment generating function
\begin{equation}\label{}
    \mathcal{M}_k(A,B) = \frac{d^k}{dt^k}M(t)\Bigr|_{t=0},\quad M(t) = e^{At + Bt^2/2},
\end{equation}
or a recurrence relation
\begin{equation}\label{}
    \mathcal{M}_{k+1}(A,B) = A\mathcal{M}_k(A,B) + B k \mathcal{M}_{k-1}(A,B)
\end{equation}
with $\mathcal{M}_0(A,B) \equiv 1$.

The process of amplification can be simplistically viewed as applying a number of annihilation operators to the density operator of a sufficiently thermalized coherent state with $\hat{\rho}_G$ given by (\ref{Pfun}) and (\ref{thermalcoh}). The amplified state is then proportional (up to normalization) to $\hat{a}^M \hat{\rho}_G \hat{a}^{\dag M}$, where $\hat{a}$ and $\hat{a}^{\dag}$ are annihilation and creation operators and $\mu$ can be easily found.
However, although this approach gives a reliable approximation for the idealized version of the tapping and measurement processes and provides a worthwhile insight into the limits of the method, the realistic procedure with a finite reflectivity of a beam splitter and inefficient photon detection is better served by the P-function treatment. In this picture, the amplified state can be expressed as
\begin{equation}\label{}
\hat{\rho}'_{\mathrm{amp}} = \frac{1}{P_S}\int
\Phi\left(\frac{\beta}{\sqrt{T}}\right)\mathcal{P}\left(\frac{\beta}{\sqrt{T}}\right)
|\beta\rangle\langle\beta| \frac{d^2\beta}{T},
\end{equation}
where $\Phi(\beta)$ is the P-function of a the coherent state with added noise (\ref{thermalcoh}), $P_S$ is the success probability and $\mathcal{P}(\beta)$ represents the probability of a successful post-selection event for a coherent state $|\beta\rangle$. For a realistic $M$ photon measurement implemented by a beam splitter with transmissivity $T$ and a photo-detector with quantum efficiency $\eta$, the $\mathcal{P}(\beta)$ can be expressed as
\begin{equation}\label{}
\mathcal{P}_{\Pi}(\beta) = \langle \sqrt{\eta(1-T)}\beta |\hat{\Pi}|\sqrt{\eta(1-T)} \beta \rangle,
\end{equation}
where we have modeled the detector inefficiency by a virtual beam splitter with transmissivity $\eta$, and $\hat{\Pi} = 1-\sum_{k=0}^{M-1} |k\rangle\langle k|$ represents the positive detection POVM element of the ideal detector. We can then obtain 
\begin{eqnarray}
  \mu = \frac{1}{P_S} \sum_{n=0}^{\infty}\frac{1}{n!\sqrt{n+1}}   \\
  \times \left[\frac{e^{-\frac{|\alpha|^2T}{\Xi_1}}}{\Xi_1}\mathcal{I}_n \left(\frac{\alpha\sqrt{T}}{\Xi_1},\frac{N_\mathrm{th}T}{2\Xi_1}\right)
  - \frac{e^{-\frac{|\alpha|^2(T+\eta(1-T))}{\Xi_2}}}{\Xi_2}\sum_{k=0}^{M-1} \frac{ \left(\eta\frac{1-T}{T}\right)^{2k}}{k!} \mathcal{I}_{n+k}\left(\frac{\alpha\sqrt{T}}{\Xi_2},\frac{N_\mathrm{th}T}{2\Xi_2}\right)\right], \nonumber
\end{eqnarray}
with $\Xi_1 = N_{th} T + 1$ and $\Xi_2 = N_\mathrm{th} T + N_\mathrm{th}\eta(1-T) + 1$. For the description to be complete we need to evaluate the success probability $P_S = \int\Phi(\beta/\sqrt{T})\mathcal{P}_{\Pi}(\beta/\sqrt{T})d^2\beta/T$. In a complete analogy with the calculations used previously we can express the success probability as
\begin{equation}\label{}
    P_S = 1-\frac{e^{-\frac{|\alpha|^2\eta(1-T)}{\Xi_3}}}{\Xi_3}\sum_{k=0}^{M-1} \frac{\left(\eta\frac{1-T}{T}\right)^{2k}}{k!}\mathcal{J}_k\left(\frac{\alpha T}{\Xi_3},\frac{N_\mathrm{th}T}{2 \Xi_3}\right),
\end{equation}
where $\Xi_3 = N_\mathrm{th}\eta(1-T)+1$ and
\begin{equation}\label{}
    \mathcal{J}_n(A,B) = \sum_{k=0}^{n}{n\choose k} \mathcal{M}_{2k}(A_r,B)\mathcal{M}_{2(n-k)}(A_i,B)
\end{equation}
is defined analogically to (\ref{Idef}).

\section* {ACKNOWLEDGEMENTS}
This work was supported by the EU project COMPAS, the BIOP Graduate school, the Lundbeck foundation and the DFG project LE 408/19-1. R.F. acknowledges support from projects No. MSM 6198959213 and No. LC06007 of the
Czech Ministry of Education, the Grant 202/08/0224 of GA CR and the Alexander von Humboldt Foundation.
P. M. acknowledges support from the Grant P205/10/P319 of GA CR.

\end{document}